\documentclass{article}
\usepackage{graphics}

\begin{document}

\title{A Qualitative Description of Boundary Layer Wind Speed Records}

\author{Rajesh G. Kavasseri \dag ~ and  Radhakrishnan Nagarajan
\ddag \\
\dag Department of Electrical and Computer Engineering \\
North Dakota State University, Fargo, ND 58105 - 5285 \\email :
rajesh.kavasseri@ndsu.edu \\
\ddag 629 Jack Stephens Drive, \# 3105 \\ University of Arkansas for
Medical Sciences, Little Rock, Arkansas 72212}

\date{}

\maketitle

{\bf keywords : wind speed, self-similarity, multifractal scaling,
atmosphere, boundary layer}

\begin{abstract}
The complexity of the atmosphere endows it with the property of
turbulence  by virtue of which, wind speed variations in the
atmospheric boundary layer (ABL) exhibit highly irregular
fluctuations that persist over a wide range of temporal and spatial
scales. Despite the large and significant body of work on microscale
turbulence, understanding the statistics of atmospheric wind speed
variations has proved to be elusive and challenging. Knowledge about
the nature of wind speed at ABL has far reaching impact on several
fields of research such as meteorology, hydrology, agriculture,
pollutant dispersion, and more importantly wind energy generation.
In the present study, temporal wind speed records from twenty eight
stations distributed through out the state of North Dakota (ND,
USA), ($\sim$ 70,000 square-miles) and spanning a period of nearly
eight years are analyzed. We show that these records exhibit a
characteristic broad multifractal spectrum irrespective of the
geographical location and topography. The rapid progression of air
masses with distinct qualitative characteristics originating from
Polar regions, Gulf of Mexico and Northern Pacific account for
irregular changes in the local weather system in ND. We hypothesize
that one of the primary reasons for the observed multifractal
structure could be the irregular recurrence and confluence of these
three air masses.
\end{abstract}

\section{Introduction}
Atmospheric phenomena are accompanied by variations at spatial and
temporal scales. In the present study, qualitative aspects of
temporal wind speed data recorded at an altitude of 10 ft from the
earth's surface are discussed. Such recordings fall under the ABL,
which is the region 1-2 km from the earths surface \cite{garratt}.
Flows in the ABL, which are generally known to be turbulent, are
influenced by a number of factors including shearing stresses,
convective instabilities, surface friction and topography
\cite{garratt, monin}. The study of laboratory scale turbulent
velocity fields has received a lot of attention in the past (see
\cite{warhaft} for a summary). A. N. Kolmogorov \cite{kolmo41a,
kolmo412}, (K41) proposed a similarity theory where energy in the
inertial sub-range is cascaded from the larger to smaller eddies
under the assumption of local isotropy. For the same, K41 statistics
is also termed as small-scale turbulence statistics. The seminal
work of Kolmogorov encouraged researchers to investigate scaling
properties of turbulent velocity fields using the concepts of
fractals \cite{mandel74}. Subsequent works of Parisi and Frisch
\cite{parisifrisch85}, Meneveau and Srinivasan, \cite{srini87_cas,
srini87a, srimulti911} provided a multifractal description of
turbulent velocity fields. While there has been a precedence of
scaling behavior in turbulence at microscopic scales \cite{kolmo41a,
kolmo412, mandel74,parisifrisch85,srini87_cas,srini87a, srimulti911,
argoul89} it is not necessary that such  a scaling manifest itself
at macroscopic scales, although there have been indications of
``unified scaling" models of atmospheric dynamics, \cite{lovejoy01}.
Several factors can significantly affect the behavior of a complex
system such as ABL \cite{monin, garratt}. Thus, an extension of
these earlier findings \cite{kolmo41a, kolmo412,
mandel74,parisifrisch85,srini87_cas,srini87a, srimulti911, argoul89}
to the present study is neither immediate, nor obvious. Attempts
have also been made to simulate the behavior of the  ABL
\cite{ding01,moeg84}. However, there are implicit assumptions made
in these studies and often, there can be significant discrepancies
between simulated flows and the actual phenomenon when these
assumptions are violated \cite{warhaft}. On the other hand,
knowledge about the nature of wind speed at ABL has far reaching
impact on several fields of research. In particular, the need to
obtain accurate statistical descriptions of flows in the ABL from
actual site recordings is both urgent and important, given its
utility in the planning, design and efficient operation of wind
turbines, \cite{peinke}. Therefore, analysis of wind speed records
based on numerical techniques is gaining importance in the recent
years. In \cite{bmschulz}, long term daily records of wind speed and
direction were represented as a two dimensional random walk and the
results reinforce the important role that memory effects have on the
dynamics of complex systems. In \cite{holgerkantz}, the short term
memory of recorded surface wind speed records is utilized to build
$m$'th order Markov chain models, from which probabilistic forecasts
of short time wind gusts are made. In \cite{govindan04}, the authors
study the correlations in wind speed data sets over a span of 24
hours, using  detrended fluctuation analysis (DFA), \cite{peng1994}
and its extension, multifractal-DFA (MF-DFA)\cite{kantel_mfdfa}.
Their studies show that the records display long range correlations
with a fluctuation exponent of $\alpha \sim 1.1$ along with a broad
multifractal spectrum. In addition, they also suggest the need for
detailed analysis of data sets from several other stations to
ascertain whether such features are characteristic of wind speed
records. In \cite{santhanam}, it is shown that rare events such as
wind gusts in wind speed data sets that are long range correlated
are themselves long range correlated. In \cite{lauren}, it is shown
that surface layer wind speed records can be characterized by
multiplicative cascade models with different scaling relations in
the microscale inertial range and the mesoscale. Our previous
studies, \cite{kgrad_ieee_dfa} suggest that at short time scales,
hourly average wind speed records are characterized by a scaling
exponent $\alpha \sim 1.4$ and at large time scales, by an exponent
of $\alpha \sim 0.7$. A deeper examination of the  data sets in
\cite{kgrad_chaos_mfdfa} using MF-DFA indicated that the records
also admitted a broad multifractal spectrum under the assumption of
a binomial multiplicative cascade model. Interestingly, scaling
phenomena have also been found in fluctuations of meteorological
variables that influence wind speed such as air humidity,
\cite{vattay}, temperature records and precipitation
\cite{abunde_kuppai}. In \cite{abunde_kuppai}, it is observed that
while temperature records display long range correlations ($\alpha
\sim 0.7$), they do not display a broad multifractal spectrum. On
the other hand, precipitation records display a very weak degree of
correlation with ($\alpha \sim 0.5$, \cite{abunde_kuppai}). While it
is difficult to directly relate the scaling results of these
variables to that of wind speed, greater insight can be gained by
analyzing data sets that are recorded over long spans from different
meteorological stations. Motivated by these findings, we chose to
investigate the temporal aspects of wind speed records dispersed
over a wide geographical area. In the present study, we follow a
systematic approach in determining the nature of the scaling of wind
speed records recorded at an altitude of 10 ft across 28 spatially
separated locations  spanning an area of approximately 70,000 sq.mi
and recorded over a period of nearly 8 years in the state of North
Dakota.

\noindent As noted earlier, convective instabilities and topography
can have a prominent impact of the flow in ABL. The air motion over
North Dakota is governed by the flow of three distinct air masses
with distinct qualities, namely from : (i) the polar regions to the
north (cold and dry) (ii) the Gulf of Mexico to the south (warm and
moist) and (iii) the northern pacific region (mild and dry). The
rapid progression and interaction of these air masses results in the
region being subject to considerable climactic variability. These in
turn can have a significant impact on the convective instabilities
which governs the flow in ABL. On the other hand, the topography of
the region has sharp contrasts on the eastern and western parts of
the state because of their approximate separation by the boundary of
continental glaciation. The eastern regions have a soft topography
compared to the western region which comprises mostly of rugged
bedrock.  In the present study, we show that the qualitative
characteristics of the wind speed records do not change across the
spatially separated locations despite the contrasting topography.
This leads us to hypothesize that the confluence of the air masses
as opposed to the topography plays a primary role in governing the
wind dynamics over ND.

\section{Methods}
\begin{figure}[htbp]
\begin{center}
\centering{\resizebox{10cm}{!}{\includegraphics{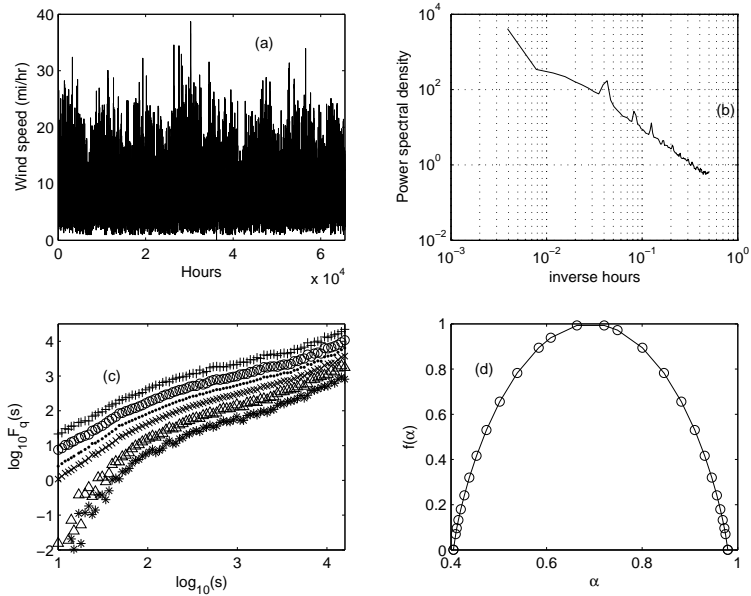}}}
\caption{(a) Temporal trace of hourly average wind speed record
(miles/hour) at one of the representative stations (Baker 1N, refer
to Table 1 for details) over a period of nearly 8 years. (b) The
corresponding power spectrum exhibits a power law decay of the form
($S(f) \sim 1/f^{\beta}$). Superimposed on the power spectrum are
prominent peaks which correspond to multiple sinusoidal trends. (c)
Log-Log plots of the fluctuation function versus time scale, $F_q(s)
\;$vs$ \;s$ for the moments $q = -10 ($*$)$, -6 (triangle ),
-0.2$(\times)$ , 2 (.), 6$(\circ)$ and $q = 10 (+)$ (d) Multifractal
spectrum of the record determined under the assumption of a binomial
multiplicative cascade model.} \label{plate1}
\end{center}
\end{figure}

\noindent Spectral analysis of stationary processes is related to
correlations in it by the Wiener-Khinchin theorem, \cite{papoulis}
and has been used for detecting possible long-range correlations. In
the present study, we observed broad-band power-law decay
superimposed with peaks. This spectral signature was consistent
across all the 28 stations (Fig. \ref{plate1}(b)). Such power-law
processes lack well-defined scales and have been attributed to
self-organized criticality, intermittency, self-similarity
\cite{bak, manneville} and multiscale randomness \cite{hausdorff}.
Superimposed on the power-law spectrum, were two high frequency
peaks which occur at $t_1 = 24$ and $t_2 = 12$ hours corresponding
to diurnal and
semi-diurnal cycles respectively. \\

\noindent Power-law decay of the power-spectrum (Fig.
\ref{plate1}(b)) can provide cues to possible long-range
correlations, but, however, it is susceptible to trends and
non-stationarities which are ubiquitous in recordings of natural
phenomena. While several estimators \cite{hurst,abryveitch98} have
been proposed in the past for determining the scaling exponents from
the given data, detrended fluctuation analysis (DFA),
\cite{peng1994} and its extension, generalized multifractal-DFA
(MF-DFA)\cite{kantel_mfdfa} have been widely used to determine the
nature of the scaling in data obtained from diverse settings
\cite{kantel_mfdfa,kantel_river,livina,havlin,yosef}. In DFA, the
scaling exponent for the given monofractal data is determined from
least-squares fit to the log-log plot of the second-order
fluctuation functions versus the time-scale, i.e. $F_q(s)\; $vs$\;s$
where $q = 2$. For MF-DFA, the variation of the fluctuation function
with time scale is determined for varying $q \;(q \ne 0$). The
superiority of DFA and MF-DFA to other estimators along with a
complete description is discussed elsewhere \cite{kantel_river}. DFA
and MF-DFA procedures follow a differencing approach that can be
useful in eliminating local trends \cite{peng1994}. However, recent
studies have indicated the susceptibility of DFA and MF-DFA to
higher order polynomial trends. Subsequently, DFA-$n$
\cite{kantelndfa} was proposed to eliminate polynomial trends up to
order $n-1$. In the present, study we have used polynomial
detrending of order four. However, such an approach might not be
sufficient to remove sinusoidal trends which can be periodic
\cite{kunhu} or quasiperiodic (see discussion in Appendix A). \\

\noindent Data sets spanning a number of years, as discussed here,
are susceptible to seasonal trends that can be periodic or
quasiperiodic in nature. Such trends manifest themselves as peaks in
the power spectrum and their power is a fraction of the broad-band
noise. These trends can also introduce spurious crossovers as
reflected by log-log plot of $F_q(s)\; $vs$\;s$  and prevent
reliable estimation of the scaling exponent. Such crossovers
indicate spurious existence of multiple scaling exponents at
different scales and shift towards higher time scales with
increasing order of polynomial detrending \cite{kantelndfa}. Thus,
it is important to discern correlations that are an outcome of
trends from those of the power-law noise. In a recent study,
\cite{svdpaper}, a singular-value decomposition (SVD) based approach
was proposed to minimize the effect of the various types of trends
superimposed on long-range correlated noise. However, SVD is a
linear transform and may be susceptible when linearity assumptions
are violated. Therefore, we provide a qualitative argument to
identify multifractality in wind speed records superimposed with
periodic and/or quasiperiodic trends. Multifractality is reflected
by a change in the slope of the log-log fluctuation plots with
varying $q$ with $(q \ne 0)$ \cite{kantel_mfdfa}. For a fixed $q$,
one observes spurious crossovers in monofractals well as
multifractal data sets superimposed sinusoidal trends. Thus
nonlinearity or a crossover of the log-log plot for a fixed $q$
might be due to trends as opposed to the existence of multiple
scaling exponents at different scales. However, we show (see
discussion under Appendix A) that the nature of log-log plot of
$F_q(s)\; $vs$\;s$ does not change with varying $q$ for monofractal
data superimposed with sinusoidal trends. However, a marked change
in the nature if the log-log plots $F_q(s)\; $vs$\;s$  with varying
$q$ is observed for multifractal data superimposed with trends.
Therefore, in the present study, the log-log plot of $F_q(s)$ vs $s$
with varying $q$ is used as a qualitative description of
multifractal structure in the given data. For the wind-speed records
across the 28 stations, we found the peaks in the power spectrum to
be consistent across all the 28 stations. Thus any effects due to
the trend on the multifractal structure, we believe would be
consistent across the 28
stations.\\

\section{Results}
\noindent MF-DFA was applied to the data sets recorded at the 28
stations. The log-log plots of the fluctuation functions ($F_q(s)\;$
vs $\;s$) with varying moments $q$ = -10, -6, -0.2, 2, 6, 10) using
fourth order polynomial detrending for one of the representative
records is shown in Fig.\ref{plate1}(c). From Fig.\ref{plate1}(c),
it can be observed that the data sets exhibit different scaling
behavior with varying $q$, characteristic of a multifractal process.
This has to be contrasted with monofractal data whose scaling
behavior is
indifferent to the choice of $q$ in the presence or absence of sinusoidal trends. \\

\begin{figure}[htbp]
\centering{\resizebox{10cm}{!}{\includegraphics{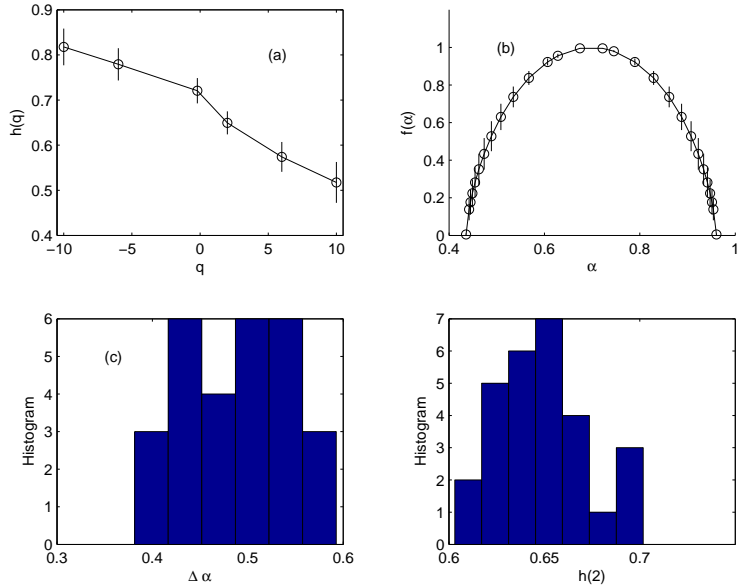}}}
\caption{(a) The mean (circle) and the standard deviation (vertical
lines) of the generalized Hurst exponent, $h(q)$ vs $q$ across the
28 stations (b) The mean (circle) and the standard deviation
(vertical lines) of the multifractal spectrum $f(\alpha)$ across the
28 stations. (c) Histogram of the multifractal widths $(\Delta
\alpha)$ across the 28 stations (d) Histogram of the Hurst exponent
$h(2)$.} \label{plate2}
\end{figure}

\noindent To compute the $q$ dependence of the scaling exponent
$h(q)$, we select the time scale in the range [2.2 - 3.7] where the
scaling was more or less constant for a given $q$. Note that this
corresponds to variations in a time span of $[10^{2.2} - 10^{3.7}]
\sim [158 - 5012]$ {\bf hours}, for a given $q$. In this range, the
slope of the fluctuation curves $h(q)$ was calculated for every $q$
and for ever station. The mean of the generalized exponents $h(q)$
over all the twenty eight stations along with the standard deviation
bars are shown in Fig. \ref{plate2}(a).

\noindent It can be noted from Fig. \ref{plate2}(a) that the slopes
$h(q)$ decrease as the moment $q$ varies from negative to positive
values which signifies that wind speed fluctuations are
heterogeneous and thus, a range of exponents is necessary to
completely describe their scaling properties. To capture this notion
of multifractality, we estimate the classical Renyi exponents
$\tau(q)$  and the singularity spectrum \cite{feder88} under the
assumption of binomial multiplicative process
\cite{barabasi,feder88,kantel_mfdfa} (see Appendix A for details).
The singularity spectrum of one of the representative stations
(Baker, 1N) is shown in Fig.\ref{plate1}(d) and its variation across
the 28 stations is shown in Fig. \ref{plate2}(b). The fitting
parameters $a,b$ for the cascade model, the Hurst exponent $h(2)$
and the multifractal width $\Delta \alpha$ for all the stations are
summarized in Table 1.

\begin{table}[htbp]
\renewcommand{\arraystretch}{1.1}
\caption{Names and locations of the 28 recording stations. The
fitting parameters $(a,b)$ of the cascade model, the Hurst exponent
$h(2)$ and the multifractal width $(\Delta \alpha)$ are also
indicted.} \label{site_labels} \centering
\begin{tabular}{|c|c|c|c|c|c|}
\hline Station Number & Name & $a$ & $b$ & $h(2)$ & $\Delta \alpha$ \\
\hline  1 & Baker 1N & 0.513 & 0.710 & 0.702 & 0.469\\
\hline  2 & Beach 9S & 0.523 & 0.76 & 0.643 & 0.539\\
\hline  3 & Bottineau 14W &0.525 & 0.720 & 0.699 & 0.456 \\
\hline  4 & Carrington 4N & 0.521 & 0.721 & 0.638 & 0.468\\
\hline  5 & Dazey 2E & 0.551 & 0.722 & 0.609 & 0.390 \\
\hline 6 & Dickinson 1NW & 0.535 & 0.715 & 0.670 & 0.418\\
\hline 7 & Edgeley 4SW & 0.531 & 0.752 & 0.573 & 0.510\\
\hline 8 & Fargo 1NW & 0.505 & 0.719 & 0.641 & 0.510 \\
\hline 9 & Forest River 7WSW & 0.556 & 0.739 & 0.670 & 0.411\\
\hline 10 & Grand Forks 3S &0.527 & 0.757 & 0.646 & 0.522\\
\hline 11 & Hazen 2W & 0.544 & 0.733 & 0.672 & 0.430\\
\hline 12 & Hettinger 1NW & 0.523 & 0.748 & 0.610 & 0.516 \\
\hline 13 & Hillsboro 7SE & 0.522 & 0.751 & 0.620 & 0.526\\
\hline 14 & Jamestown 10 W & 0.547 & 0.739 & 0.636 & 0.434\\
\hline 15 & Langdon 1E & 0.545 & 0.710 & 0.644 & 0.381\\
\hline 16 & Linton 5N & 0.531 & 0.711 & 0.623 & 0.422\\
\hline 17 & Minot 4S & 0.550 & 0.706 & 0.629 & 0.459 \\
\hline 18 & Oakes 4S & 0.507 & 0.755 & 0.657 & 0.574\\
\hline 19 & Prosper 5NW & 0.504 & 0.720 & 0.671 & 0.513\\
\hline 20 & Mohall 1W & 0.513 & 0.753 & 0.654 & 0.554\\
\hline 21 & Streeter 6NW & 0.521 & 0.742 & 0.675 & 0.509 \\
\hline 22 & Turtle Lake 4N & 0.546 & 0.729 & 0.693 & 0.418\\
\hline 23 & Watford City 1W & 0.524 & 0.768 & 0.631 & 0.551\\
\hline 24 & St. Thomas 2WSW & 0.550 & 0.744 & 0.633 & 0.434\\
\hline 25 & Sidney 1NW & 0.518 & 0.772 & 0.635 & 0.593\\
\hline 26 & Cavalier 5W & 0.506 & 0.736 & 0.651 & 0.539\\
\hline 27 & Williston 5SW & 0.514 & 0.759 & 0.635 & 0.562 \\
\hline 28 & Robinson 3NNW & 0.514 & 0.739 & 0.620 & 0.525\\ \hline
\end{tabular}
\end{table}

\begin{figure}[htbp]
\centering{\resizebox{10cm}{!}{\includegraphics{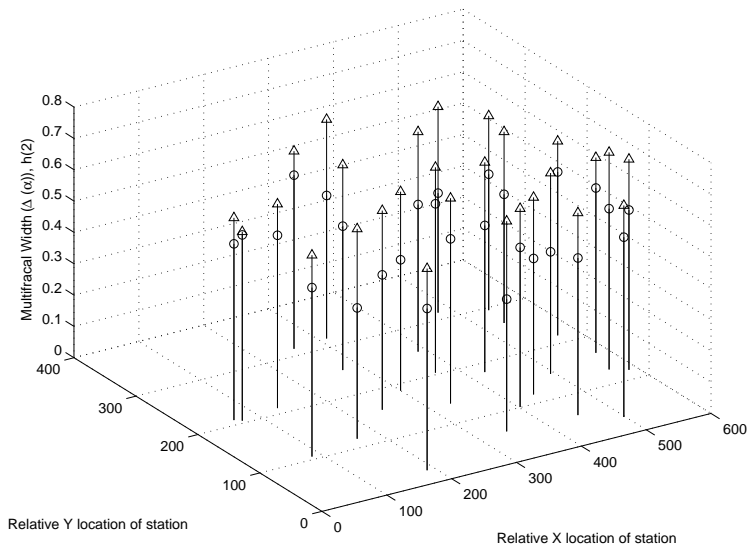}}}
\caption{The multifractal width for each of the 28 stations is
indicated by circles. The Hurst exponent $h(2)$ is indicated by
upright triangles. The x-y plane represents the x and y coordinate
in the of North Dakota.} \label{mfwidths}
\end{figure}

\noindent These results indicate multifractal scaling consistent
across the stations. Earlier studies
\cite{mandel68,ivanov,kantel_mfdfa} have suggested the choice of
random shuffled surrogates to rule out the possibility that the
observed fractal structure is due to correlations as opposed to
broad probability distribution function. The wind speeds in the
present study follow a two-parameter asymmetric Weibull distribution
whose parameters were also similar across the 28 stations. MF-DFA on
the random shuffle surrogates of the original records
Fig.\ref{plate2}(d) indicate a scaling of the form $F_q(s) \sim
s^{0.5}$ with varying $q$, characteristic of random noise and loss
of multifractal structure. The width of the multifractal spectrum
 was used to characterize the strength of multifractality. The
histogram of the multifractal widths obtained across the 28
stations, Fig. \ref{plate2}(c), was narrow with mean and standard
deviation $\Delta (\alpha) = (0.4866 \pm 0.0599)$. The multifractal
widths and the Hurst exponent $h(2)$ across the twenty eight
stations is also shown in Fig.\ref{mfwidths}.

\noindent In the present study a systematic approach was used to
determine possible scaling in the temporal wind-speed records over
28 spatially separated stations in the state of North Dakota.
Despite the spatial expanse and contrasting topography the multifractal qualitative
characteristics of the wind speed records as reflected by singularity spectrum, were
found to be similar. Thus multifractality may be an invariant feature in describing the
dynamics long-term motion of wind speed records in ABL over North Dakota. We also
believe that the irregular recurrence and confluence of the air masses from Polar,
Gulf of Mexico and the northern Pacific may play an important role in explaining the
 observed multifractal structure.

\section*{Acknowledgments} The financial support from
ND-EPSCOR through NSF grant EPS 0132289 is gratefully
acknowledged.\\

\newpage

\appendix

\section{Data Acquisition, MF-DFA algorithm and Discussion}

\subsection{Data Acquisition}
The wind speed records at the 28 stations spanning nearly 8 years
were obtained from part of the climatological archives of the state
of North Dakota. Stations were selected to represent the general
climate of the surrounding area. Wind speeds were recorded by means
of conventional cup type anemometers located at a height of 10 ft.
The anemometers have a range of 0 to 100 mph with an accuracy of
$\pm 0.25$ mph. Wind speeds acquired every five seconds are averaged
over a 10 minute interval to compute the 10 minute average wind
speed. The 10 minute average wind speeds are further averaged
over a period of one hour to obtain the hourly average wind speed.\\

\subsection{Multifractal Detrended Fluctuation Analysis (MF-DFA):}
MF-DFA, \cite{kantel_mfdfa} a generalization of DFA has been shown
to reliably extract more than one scaling exponent from a time
series. A brief description of the algorithm is provided here for
completeness. A detailed explanation can be found elsewhere
\cite{kantel_mfdfa}. Consider a time series $\{x_k\}, k = 1 \dots
N$. The MF-DFA algorithm consists of the following steps.
\begin{enumerate}
\item The series $\{x_k\}$ is integrated to form the integrated
series $\{y_k\}$ given by
\begin{equation}
y(k) = \sum_{i=1}^{i=k} [x(i) - \bar{x}] \;\;\; k = 1, \dots N
\end{equation}
where $\bar{x}$ represents the average value.

\item The series $\{y_k\}$ is divided in to $n_s$ non-overlapping
boxes of equal length$s$ where $n_s = int(N/s)$. To accommodate the
fact that some of the data points may be left out, the procedure is
repeated from the other end of the data set \cite{kantel_mfdfa}.

\item The local polynomial trend $y_v$ with order $v$ is fit to the data in each box,
the corresponding variance is given by

\begin{equation} F^2(v,s) = \left \{ \frac{1}{s} \sum_{i=1}^{i=s}
\{y [N-(v-n_s)s + i] - y_v(i) \right \}^2
\end{equation}

for $v = 1, \dots n_s$. Polynomial detrending of order m is capable
of eliminating trends up to order m-1. \cite{kantel_mfdfa}

\item The $qth$ order fluctuation function is calculated from
averaging over all segments.

\begin{equation}
F_q(s) = \left \{ \frac{1}{2 n_s} \sum_{i=1}^{i= 2
n_s}[F^2(v,s)]^{q/2}  \right \} ^{1/q}
\end{equation}
In general, the index $q$ can take any real value except zero.

\item Step 3 is repeated over various time
scales $s$. The scaling of the fluctuation functions $F_q(s)$ versus
the time scale $s$ is revealed by the log-log plot.

\item The scaling behavior of the fluctuation functions is
determined by analyzing the log-log plots $F_q(s)$ versus $s$ for
{\em each} $q$. If the original series $\{x_k \}$ is power-law
correlated, the fluctuation function will vary as
\begin{equation}
F_q(s) \sim s^{h(q)} \label{hq}
\end{equation}
\end{enumerate}

\noindent The MF-DFA algorithm \cite{kantel_mfdfa} was used
 to compute the multifractal fluctuation
functions. The slopes of the fluctuation functions ($h(q)$) for each
\newline  $q = (-10, -6, -0.2, 2, 6, 10)$ was estimated by linear
regression over the time scale range $s = [2.2, 3.7]$. The
generalized Hurst exponents ($h(q)$) are related to the Renyi
exponents $\tau(q)$ by $q h(q) = \tau(q) + 1 $. The multifractal
spectrum $f(\alpha_{h})$ defined by, \cite{feder88} $\alpha_{h} =
\frac{d \tau(q)}{d q},\;\; f(\alpha_{h}) = q \alpha_{h} - \tau(q)$.
Under the assumption of a binomial multiplicative cascade model
\cite{feder88} the generalized exponents $h(q)$ can be determined
from $h(q) = \frac{1}{q} - \frac{ln(a^q + b^q)}{q ln2}
\label{mfcascade} $. The parameters $a$ and $b$ for each station was
determined using a nonlinear least squares fit of the preceding
formula with those calculated numerically. Finally, the multifractal
width was calculated using $\Delta \alpha = \Delta \alpha_{h} =
h(-\infty) - h(\infty) =
\frac{(ln(b) - ln(a))}{ln2}$, \cite{kantel_mfdfa}.\\

\subsection{Discussion}

\noindent The power spectrum of the wind speed records considered in
the present study exhibited a power law decay of the form ($S(f)
\sim 1/f^{\beta}$) superimposed with prominent peaks which
corresponds to multiple sinusoidal trends. Such a behavior is to
 expected on data sets spanning several years. The nature of the power spectral signature
 was consistent across all stations. This enables us to compare the nature
of scaling across the 28 stations. Recent studies had indicated the
susceptibility of  MF-DFA to sinusoidal trends in the given data,
\cite{kunhu}. Sinusoidal trends can
 give rise to spurious crossovers and nonlinearity in the log-log plot that indicate
 the existence of more than one scaling exponent.  In a recent study \cite{svdpaper}, we had successfully
 used the singular-value decomposition to minimize the effect of offset, power-law,
 periodic and quasi-periodic trends.
However, SVD is a linear transform and may be susceptible when
linearity assumptions are violated. Estimating the SVD for
large-embedding matrices is computationally challenging. Therefore,
in the present study we opted for a qualitative description of
multifractal structure by inspecting the nature of the log-log plots
of the fluctuation function versus time scale $F_q(s) \;$vs $\;s$
with varying moments $q$. We show that the nature of the log-log
plot does not show appreciable change with varying moments $q$ for
monofractal data superimposed with sinusoidal trends. However, a
marked change in the nature of the log-log plot is observed for
multifractal data superimposed with sinusoidal trends. Moreover, the
nature of the trends as reflected by the power spectrum is
consistent across the 28 stations. This enables us to compare the
multifractal description obtained across the stations. The
effectiveness of the qualitative description is demonstrated with
synthetic monofractal and multifractal data sets superimposed with
sinusoidal trends.

\subsubsection{MF-DFA results of monofractal and multifractal data superimposed with sinusoidal
trends}

\begin{figure}[htbp]
\centering{\resizebox{8.0cm}{!}{\includegraphics{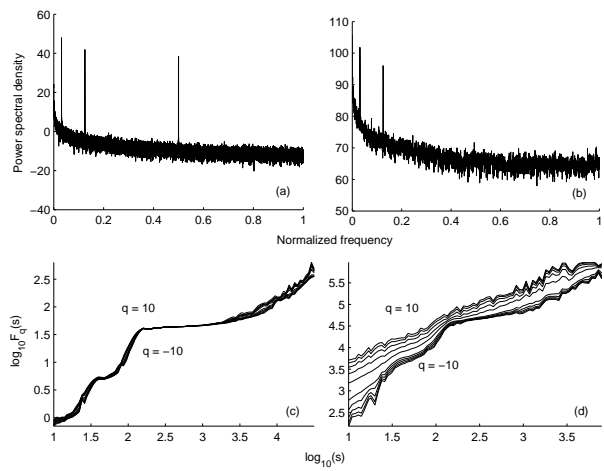}}}
\caption{MF-DFA studies of monofractal and multifractal data sets
superimposed with multiple sinusoidal trends.} \label{trndargu}
\end{figure}

Consider a signal $y_1(n)$ consisting of monofractal data $s_1(n)$
superimposed with a sinusoidal trend $t_1(n)$. Let $y_2(n)$ be a
signal consisting of multifractal data $s_2(n)$ superimposed with
sinusoidal trend $t_2(n)$. The trends are described by,
\begin{eqnarray*}
t_1(n) = A_1\sin(2\pi n/T_1) + A_2\sin(2\pi n/T_2) + A_3\sin(2\pi
n/T_3), n = 1 \dots N_1 \\
t_2(n) =  B_1\sin(2\pi n/T_{1b}) + B_2\sin(2\pi n/T_{2b}), n = 1
\dots N_2
\end{eqnarray*}
The signals are given by, $y_i(n) = s_i(n) + t_i(n), i = 1, 2$ where
$s_1(n)$ is monofractal data with $\alpha = 0.9$ and $s_2(n)$ is
multifractal data is that of internet log traffic, \cite{vehel}. The
dominant spectral peaks Fig.\ref{trndargu}(a) and
Fig.\ref{trndargu}(b) reflect the presence of these trends in
signals $y_1$ and $y_2$ respectively. The MF-DFA plots $F_q(s)\;$ vs
$\;s$ with fourth order detrending and $q = -10, -8, -6, -4, -2, 2,
4, 6, 8, 10$ for signals $y_1$ and $y_2$ are shown in
Fig.\ref{trndargu}(c) and Fig.\ref{trndargu}(d) respectively. For
the monofractal data, the trends introduce spurious crossover at $s
\sim 2.2$ in the log-log plot of the $F_q(s) \;$vs $\;s$ for a given
$q$. However, the nature of the log-log plots fail to show
appreciable change with varying $q$, Fig.\ref{trndargu}(c). For
multifractal data with trends, spurious crossovers are still noted
at $s \sim 2.2$ in the log-log plot of the $F_q(s) \;$vs $\;s$ for a
given $q$. However, in this case, Fig.\ref{trndargu}(d), the nature
of the log-log plots show a significant change with varying $q$
indicating multifractal scaling in the given data, unlike the case
with monofractal data with trends.

\noindent {\bf Parameters:} \newline

\noindent $A_1 = 6,\; A_2 = 3,\; A_3 = 2,\; T_1 = 2^6,\; T_2 =
2^4,\; T_3 = 2^2, N_1 = 2^{17},\; B_1 = 6000,\; B_2 = 3000, \;
T_{1b} = 2^6,\; T_{2b} =
2^4, N_2 = 2^{15}$. \\

\noindent The data sets are available from \\
http://www.physionet.org/physiobank/database/synthetic/tns/.

\end{document}